\documentstyle[12pt]{article}
\input psfig.sty
\input epsfig.sty
\def\nn{\noindent}
\def\Re{{\cal R \mskip-4mu \lower.1ex \hbox{\it e}\,}}
\def\Im{{\cal I \mskip-5mu \lower.1ex \hbox{\it m}\,}}
\def\ie{{\it i.e.}}
\def\eg{{\it e.g.}}

\def\etal{{\it et al.}}

\def\sub#1{_{\lower.25ex\hbox{$\scriptstyle#1$}}}
\def\tev{\,{\ifmmode\mathrm {TeV}\else TeV\fi}}
\def\gev{\,{\ifmmode\mathrm {GeV}\else GeV\fi}}
\def\mev{\,{\ifmmode\mathrm {MeV}\else MeV\fi}}
\def\mpl{\ifmmode \overline M_{Pl}\else $\overline M_{Pl}$\fi}
\def\to{\rightarrow}
\def\slash{\not\!}
\def\subw{_{\rm w}}
\def\mh{\ifmmode m\sbl H \else $m\sbl H$\fi}
\def\mch{\ifmmode m_{H^\pm} \else $m_{H^\pm}$\fi}
\def\mt{\ifmmode m_t\else $m_t$\fi}
\def\mc{\ifmmode m_c\else $m_c$\fi}
\def\mz{\ifmmode M_Z\else $M_Z$\fi}
\def\mw{\ifmmode M_W\else $M_W$\fi}
\def\mws{\ifmmode M_W^2 \else $M_W^2$\fi}
\def\mhs{\ifmmode m_H^2 \else $m_H^2$\fi}   
\def\mzs{\ifmmode M_Z^2 \else $M_Z^2$\fi}
\def\mts{\ifmmode m_t^2 \else $m_t^2$\fi}
\def\mcs{\ifmmode m_c^2 \else $m_c^2$\fi}
\def\mchs{\ifmmode m_{H^\pm}^2 \else $m_{H^\pm}^2$\fi}
\def\ztwo{\ifmmode Z_2\else $Z_2$\fi}
\def\zone{\ifmmode Z_1\else $Z_1$\fi}
\def\mtwo{\ifmmode M_2\else $M_2$\fi}
\def\mone{\ifmmode M_1\else $M_1$\fi}
\def\tb{\ifmmode \tan\beta \else $\tan\beta$\fi}
\def\xw{\ifmmode x\subw\else $x\subw$\fi}
\def\ch{\ifmmode H^\pm \else $H^\pm$\fi}
\def\lum{\ifmmode {\cal L}\else ${\cal L}$\fi}
\def\inpb{\,{\ifmmode {\mathrm {pb}}^{-1}\else ${\mathrm {pb}}^{-1}$\fi}}
\def\infb{\,{\ifmmode {\mathrm {fb}}^{-1}\else ${\mathrm {fb}}^{-1}$\fi}}
\def\epem{\ifmmode e^+e^-\else $e^+e^-$\fi}
\def\ppb{\ifmmode \bar pp\else $\bar pp$\fi}
\def\bsg{\ifmmode B\to X_s\gamma\else $B\to X_s\gamma$\fi}
\def\bsll{\ifmmode B\to X_s\ell^+\ell^-\else $B\to X_s\ell^+\ell^-$\fi}
\def\bstt{\ifmmode B\to X_s\tau^+\tau^-\else $B\to X_s\tau^+\tau^-$\fi}
\def\lamt{\ifmmode \tilde\lambda\else $\tilde\lambda$\fi}
\def\shat{\ifmmode \hat s\else $\hat s$\fi}
\def\that{\ifmmode \hat t\else $\hat t$\fi}
\def\uhat{\ifmmode \hat u\else $\hat u$\fi}

\newskip\zatskip \zatskip=0pt plus0pt minus0pt
\def\matth{\mathsurround=0pt}

\def\gsim{\mathrel{\mathpalette\atversim>}}
\def\atversim#1#2{\lower0.7ex\vbox{\baselineskip\zatskip\lineskip\zatskip
  \lineskiplimit 0pt\ialign{$\matth#1\hfil##\hfil$\crcr#2\crcr\sim\crcr}}}

\renewcommand{\thefootnote}{\fnsymbol{footnote}}

\begin{document} \begin{titlepage} 
\rightline{\vbox{\halign{&#\hfil\cr
&SLAC-PUB-8863\cr
&June 2001\cr}}}
\begin{center}

{\Large\bf Probes Of Universal Extra Dimensions at Colliders}
\footnote{Work supported by the Department of 
Energy, Contract DE-AC03-76SF00515}
\medskip

\normalsize 
{\bf \large Thomas G. Rizzo}
\vskip .3cm
Stanford Linear Accelerator Center \\
Stanford University \\
Stanford CA 94309, USA\\
\vskip .2cm

\end{center}

\begin{abstract} 
In the Universal Extra Dimensions model of Appelquist, Cheng and Dobrescu, all 
of the Standard Model fields are placed in the bulk and thus have 
Kaluza-Klein(KK) excitations. These KK states can only be pair produced at 
colliders due to the tree-level conservation of KK number, with the lightest 
of them being stable and possibly having a mass as low as $\simeq 350-400$ GeV. 
After calculating the contribution to $g-2$ in this model we investigate the 
production cross sections and signatures for these particles at both hadron 
and lepton colliders. We demonstrate that these signatures critically depend 
upon whether the lightest KK states remain stable or are allowed to decay by 
any of a number of new physics mechanisms. These mechanisms which induce KK 
decays are studied in detail. 
\end{abstract} 




\renewcommand{\thefootnote}{\arabic{footnote}} \end{titlepage}


\section{Introduction}

The possibility that the gauge bosons of the Standard Model(SM) may be 
sensitive to the existence of extra dimensions near the TeV scale has been 
known for some time{\cite{antoniadis}}. However, one finds that 
the phenomenology of these 
models is particularly sensitive to the manner in which the SM fermions (and 
Higgs bosons) are treated. 

In the simplest scenario, the fermions remain on 
the wall located at the fixed point $y_i=0$ and are not free to experience 
the extra dimensions. (Here, lower-case Roman indices 
label the co-ordinates of the additional dimensions while Greek indices 
label our usual 4-d space-time.) However, since 5-d translational invariance is 
broken by the wall, the SM fermions interact with the Kaluza-Klein(KK) 
tower excitations of the SM gauge fields in the usual trilinear manner, \ie, 
$\sim gC_n\bar f \gamma_\mu f G^\mu_{(n)}$, with the $C_n$ being some 
geometric factor and $n$ labelling the KK tower state with which the fermion 
is interacting. Current low-energy constraints arising 
from, \eg, $Z$-pole data, the $W$ boson mass and $\mu$-decay generally require 
the mass of the lightest KK gauge boson to be rather heavy,  $\gsim 4$ TeV in 
the case of the 5-d SM{\cite {bunch}} independent of whether or not the Higgs 
fields are on the wall under the assumption that 
the $C_n$ are $n$-independent for $n \geq 1$. 

A second possibility occurs when the SM fermions experience 
extra dimensions by being `stuck', \ie, localized or trapped at different 
specific points in a thick brane{\cite {nam}} away from the conventional fixed 
points. It has been shown that such a scenario can explain the absence of a 
number of rare processes, such as proton decay, by geometrically 
suppressing the size of the Yukawa couplings associated with the relevant 
higher dimensional operators without resorting to the existence of additional 
symmetries of any kind. In addition such a scenario may be able to explain the 
fermion mass hierarchy and the observed CKM mixing structure thus addressing 
important issues in flavor physics{\cite {mirsch}}. The couplings of the SM 
fermions to the gauge KK towers are in this case dependent upon their location 
in the extra dimension(s). 

A last 
possibility, perhaps the most democratic, requires all of the SM fields to 
propagate in the $\sim$ TeV$^{-1}$ bulk{\cite {ACD}}, \ie, Universal Extra 
Dimensions(UED). In this case, there being no matter on the 
walls, the conservation of momentum in the extra dimensions is restored and 
one now obtains interactions in the 4-d Lagrangian of the form 
$\sim gC_{ijk}\bar f^{(i)} \gamma_\mu f^{(j)} G^\mu_{(k)}$, which for 
flat space metrics vanishes unless $i+j+k=0$, as a result of the afore 
mentioned momentum conservation. Although this momentum conservation is 
actually broken by 
orbifolding, one finds, at tree level, that KK number remains a 
conserved quantity. (As we will discuss below this conservation law is itself  
further 
broken at one loop order.) This implies that pairs of 
zero-mode fermions, which we identify with those of the SM, cannot directly 
interact singly with any of the excited modes in the gauge boson KK towers.  
Such a situation  
clearly limits any constraints arising from precision measurements since zero 
mode fermion fields can only interact with pairs of tower gauge boson fields. 
In addition, at colliders it now follows that 
KK states must be pair produced, thus significantly 
reducing the possible direct search reaches for these states. In fact, 
employing constraints from current experimental data, Appelquist, Cheng and 
Dobrescu(ACD){\cite {ACD}} find that the KK states in this 
scenario can be as light as $\simeq 350-400$ GeV, much closer to current 
energies than the KK modes in the first case discussed above. If these states 
are, in fact, nearby, they will be copiously produced at the LHC, and possibly 
also at the Tevatron, in a variety 
of different channels. It is the purpose of this paper to estimate the 
production rates for pairs of these particles in various channels and to 
discuss their possible production signatures. This is made somewhat difficult 
by the apparent conservation of KK number which appears to forbid the decay of 
heavier excitations into lighter ones and is a point we will return to in 
detail below.

The outline of this paper is as follows: in Section II we briefly discuss the 
particle spectrum in the UED model and the breaking of KK number to `KK 
parity' at one loop. 
In Section III, in an attempt to get a further handle on the compactification 
scale of the UED scenario, we discuss the shift in the value of $g-2$ 
predicted in this model. Unfortunately, as we will see, no new constraints are 
obtained. In 
Section IV we discuss the production mechanisms and cross sections for pairs 
of KK excitations of the SM fields at the Tevatron and LHC. Similar production 
mechanisms and $e^+e^-$ and $\gamma \gamma$ colliders are also briefly 
considered. Section V 
discusses the possible signatures for KK pair production addressing the 
issue of their possible stability in light of our earlier discussions in 
Section II. Three particular decay scenarios are considered. 
A discussion and our conclusions can be found in Section VI.

\section{Model Set-up Review}

In this section we very briefly review the basic nature of the UED model. The 
essential 
idea is that all of the fields of the SM are put in the bulk and thus have 
KK excitations. For simplicity in what follows we will limit our discussion 
to the case of one extra dimension with the extension to more than one 
dimension being reasonably straightforward. Due to $S_1/Z_2$ orbifolding, 
which is necessary to obtain chiral zero mode fermions,  the 
fields can be classified as either $Z_2$ even or odd: all Higgs boson and 
4-d gauge fields are taken as $Z_2$ even whereas the 5-d components of the 
gauge fields (which are not present in the Unitary Gauge) must then be 
$Z_2$ odd. Taking the compactification radius to be $R=1/M_c$ the corresponding 
eigenfunctions are simply 
$\sim \cos ~ny/R~(\sin ~ny/R)$ for the $Z_2$ even(odd) 
fields. The gluon and photon excitations have the usual KK masses, $nM_c$, 
while the $W$ and $Z$ towers have shifted masses $[M_{W,Z}^2+(nM_c)^2]^{1/2}$ 
after spontaneous symmetry breaking(SSB) by the zero mode Higgs field. Although 
the zero mode Goldstone bosons are eaten as part of the SSB mechanism, their 
tower states remain physical, level by level 
degenerate with the gauge bosons and are also $Z_2$ even. In the 
fermion sector there is a well-known doubling of states; every $SU(2)_L$ 
doublet $D$ or singlet $S$ field has a vector-like tower of states above the 
chiral zero mode. $S_1/Z_2$ only allows for the existence of 
the left-handed (right-handed) 
zero mode for the $D$($S$). Note that while one of the fermion KK towers, the 
one matching the chirality of the zero mode, is constrained to be $Z_2$ even, 
the other must be $Z_2$ odd.) Note that in performing calculations one must 
be careful not to confuse the $S_L$ and $D_L$ fields and their 
corresponding right-handed partners. 
As will be discussed below, the zero mode Higgs vev links the $S$ and $D$ 
states level by level and simultaneously generates the zero mode fermion 
masses as usual. The Yukawa coupling of this interaction is completely fixed 
by the SM fermion mass. This cross linking of the the two towers $D$ and $S$ 
will be necessary in order to generate the $g-2$ of the muon in this model.

Now although the KK number is conserved at the tree level it becomes apparent 
that it is no longer so at loop order{\cite {strumia}}. Consider a 
self-energy diagram  with a field that has KK number of $2n(2n+1)$ entering 
and a zero($n=1$) mode leaving 
the graph; KK number conservation clearly does not forbid such an amplitude 
and constrains the 2 particles in the intermediate state to both have KK number 
$n$($n$ and $n+1$). The existence of such amplitudes implies that all even 
and odd KK states mix separately so that the even KK excitations can 
clearly decay to zero modes while odd KK states can now 
decay down to the KK number=1 state. Thus it is KK {\it parity}, $(-1)^n$, 
which remains conserved while KK number 
itself is broken at one loop. Since the lightest KK excited states with 
$n=1$ have odd KK parity they remain stable unless new physics is introduced. 
As we are only concerned with the production of pairs of the lightest KK 
particles in our discussion below, we are faced with the possibility 
of producing heavy stable states at colliders. This point will be 
discussed in detail further below.

\section{$g-2$}

The bound on $M_c$ obtained by ACD in this model is quite low and is not 
improved by the consideration of other processes such as $b\to s\gamma$ as 
discussed by Agashe, Deshpande and Wu{\cite {desh1}}. 
To see if we can improve this bound on $M_c$, we briefly 
discuss the contribution to $g-2$ in the UED model; we follow the analysis 
as given in{\cite {rsgm2,desh2}}. To this end we 
consider the specific situation where we have two fermions in the bulk, 
$D_\mu$ and $S_\mu$, corresponding to the 5-dimensional muon fields, having 
the quantum numbers of an $SU(2)_L$ doublet and singlet with weak 
hypercharges $Y=-1/2$ and $-1$, respectively. (We will drop the $\mu$ index on 
these fields in what follows as it is clearly understood what fields we are 
discussing.) The interactions of these fermions with 
the gauge fields can be described by the action, 
\begin{equation}
S_{fV} =\int d^4x \int dy \left[V^M_n\left
({i\over {2}}\overline{S} \, \gamma^n\,  {\cal D}_M S + h. c.\right) - 
sgn(y) m_S \overline{S} S +(S\to D\right)],
\label{SfV}
\end{equation}
where, $V$ is the vielbein, which is trivial for the flat space case we are 
considering, ${\cal D}_M$ is a covariant derivative and $h. c.$ 
denotes the Hermitian conjugate term. Note that gauge interactions 
do not mix the $D$ and $S$ fields.  The $D$ and $S$ fields also interact 
with the bulk Higgs isodoublet field, $H^0$, \ie, 
\begin{equation}
S_{fH} ={(2\pi R)^{1/2} \lambda }\int d^4x \int dy \ \overline{S}DH^0 \ 
+ h. c.,
\label{SfH}
\end{equation}
with $R$ being the compactification radius as described above and 
$\lambda$ being a dimensionless Yukawa coupling. 
Due to the KK mechanism the fields $D_{L,R}^{(n)}$ and $S_{L,R}^{(n)}$ 
form separate 4-d towers of Dirac fermions which, as discussed above, 
are degenerate level by level. These KK expansions can be written as 
$D=\sum D_L^{(n)}(x)\chi^{(n)}(y)+ D_R^{(n)}(x)\tau^{(n)}(y)$ and 
$S=\sum S_L^{(n)}(x)\tau^{(n)}(y)+ S_R^{(n)}(x)\chi^{(n)}(y)$ where the 
$\chi(\tau)$ fields are $Z_2$ even(odd). Note that the $Z_2$ orbifold 
symmetry and orthonormality only allows `level-diagonal' couplings of both 
types:  $\overline{D_L}^{(n)}S_R^{(n)}+h. c.$ and 
$\overline{D_R}^{(n)}S_L^{(n)}+h. c.$. The value of $\lambda$ is fixed by 
requiring the zero mode fermion  obtain a mass $m_\mu$ after the Higgs zero 
mode obtains a vev, $v$, and tells us the level by level coupling between the 
tower members. 

In terms of the $D$ 
and $S$ fields, the operator which generates the anomalous magnetic dipole 
moment of the $\mu$ can be written as $D_L^{(0)}\sigma_{\mu\nu}q^\nu S_R^{(0)}
+h.c.$. This reminds us that this operator and the muon mass generating term 
have the same isospin and helicity structure such that a Higgs interaction is 
required in the form of a mass insertion to connect the two otherwise 
decoupled zero 
modes. We can think of this mass insertion as the interaction of a fermion 
with an external Higgs field that has been replaced by its vev.

\vspace*{-0.5cm}
\nn
\begin{figure}[htbp]
\centerline{
\psfig{figure=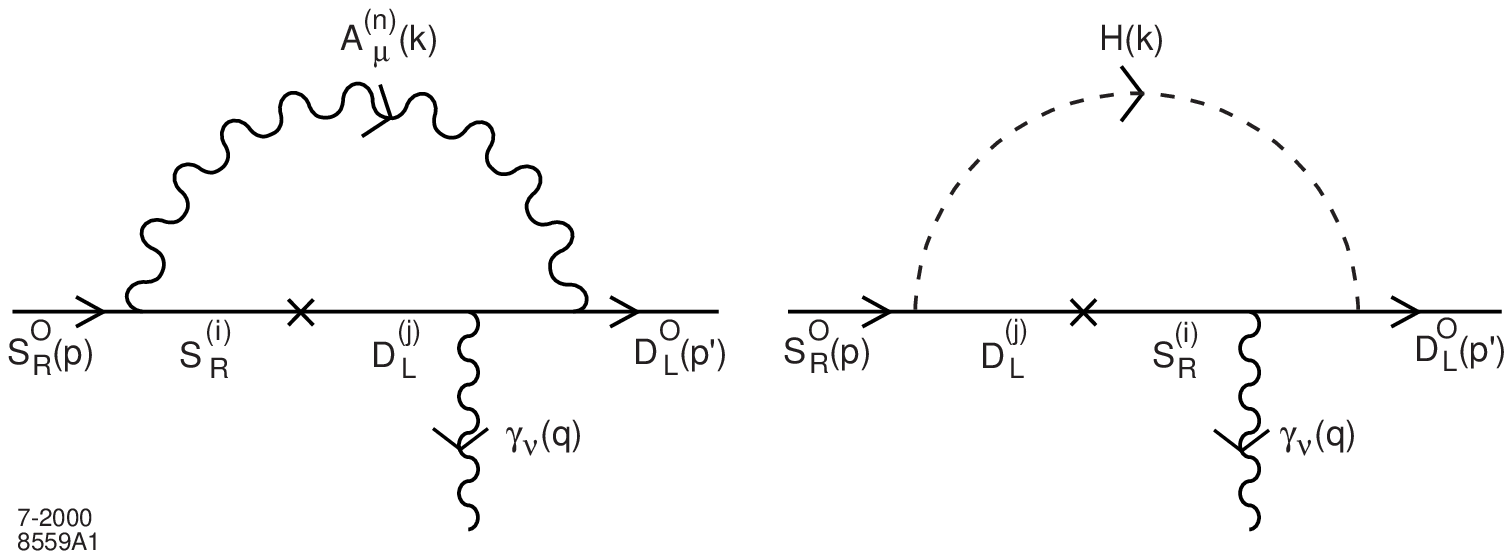,height=8.0cm,width=16.2cm,angle=0}}
\vspace*{0.1cm}
\caption[*]{Typical Feynman diagrams for the gauge and Higgs boson 
contribution to $(g-2)_\mu$. The mass insertion is denoted by the cross.}
\label{fig1}
\end{figure}
\vspace*{0.7mm}

Helicity flips play an 
important role in evaluating the contributions to $(g-2)_\mu$  
since muon KK excitations are now propagating inside the loop. 
As is well-known, for non-chiral couplings the contribution to the 
anomalous magnetic moment of a light fermion can be enhanced when a heavy 
fermion of mass $m_h$ participates inside the loop{\cite {stan}}.
There are a number of diagrams that can contribute to $(g-2)_\mu$ at one loop 
of which two are shown in Fig.~1. The diagram on the left corresponds to the 
exchange of a tower of the 4-d neutral gauge bosons, $\gamma^{(n)}$ and/or 
$Z^{(n)}$, which we will now discuss in detail. Due to gauge 
invariance we are free to choose a
particular gauge in order to simplify the calculation. Here, we make use of 
the $\xi=1$ unitary gauge where the numerator of the 
4-d propagator is just the negative of the 
flat space metric tensor{\cite {dpq}}. 
Hence, in this gauge, the loops with the 4-d components of the 
gauge fields and the ones with the fifth component need to be considered 
separately. In this example, the mass insertion takes place inside the loop 
before the photon is emitted. Clearly there are three other diagrams of this 
class: two with the mass insertion on an external leg and the third with the 
mass insertion inside the loop but after the photon is emitted. The amplitude 
arising from this vector exchange graph is given by 
\begin{eqnarray}
{\cal M}_V &=& g_Lg_R~\bar u(p')\ (ie\gamma_\mu)P_L 
~i{{\hat {\slash p'}+m_n} \over {\hat p'^2-m_n^2}} \ (-ie\gamma_\nu)P_L \, 
\nonumber \\  
 &\times &  i{{\hat {\slash p}+m_n} \over {\hat p^2-m_n^2}} \ (im_\mu P_R) 
\ ~i{{\hat {\slash p}+m_n} \over {\hat p^2-m_n^2}} \ (ie\gamma^\mu)P_R \ u(p) 
\ {-i\over {k^2-m_A^{(n)~2}}} +h. c. \,,
\end{eqnarray}
where $g_{L,R}$ are the corresponding couplings of the SM gauge boson 
to the $\mu$ in units of $e$ and $\hat p(\hat p')=p(p')-k$. 
Here (in the limit that we can neglect 
the muon mass) $m_n=nM_c$ are the masses of the $D$ or $S$ 
muonic KK states and $m_A^{(n)}$ are the corresponding 
masses of the KK gauge tower states, 
$[m_A^{(0)2}+(nM_c)^2]^{1/2}$, with $M_c$ being the compactification scale.
Note that the mass insertion, $m_\mu$, comes with a chirality factor that can 
be determined from the action $S_{fH}$. 
The amplitude where the mass insertion comes after the photon emission can 
be easily obtained by interchanging the ordering in the resulting final 
amplitude expression. 

What happens when the mass insertion occurs on the external legs? 
With some algebra it is straightforward to 
show that the corresponding amplitudes obtained in these two cases are 
suppressed in comparison to the case of internal insertion 
by at least a factor of order $\sim m_\mu/M_{KK}$, 
where $M_{KK}$ is a typical large KK mass. 
In the case of the $W$ gauge boson 
tower graphs, since the $W$ couples only to the $D$'s, 
the mass insertion must occur on the incoming leg of the graph and the photon 
is then emitted from the $W$; this graph can also 
be shown to produce a sub-leading contribution by at least a factor of 
order $\sim m_\mu/M_{KK}$. 
Thus, $W$ tower graphs can be safely ignored in comparison to those arising 
from the $Z$ and $\gamma$ towers. (Note that the suppressions that we obtain 
here in the case of heavy internal KK states are {\it absent} in the SM 
calculation of $g-2$ since the muon or its neutrino are now the internal loop 
fermion). In a similar fashion it is clear that the graphs containing 
Goldstone bosons will also be suppressed since their couplings are of order 
$m_\mu/M_W$. 

The next class of graphs is similar to the 4-d vector exchange, but in the 
$\xi=1$ gauge, now involves the fifth component of the original 5-d field. 
Here it is important to recall that these fifth components are $Z_2$ odd 
fields thus connecting $S_L(D_L)$ with $S_R(D_R)$. Let us first consider the 
case where the neutral 4-d vector 
field is replaced by the 5-d scalar field; in analogy 
with Fig.1 we obtain:
\begin{eqnarray}
{\cal M}_5 &=& g_Lg_R~\bar u(p')\ ie P_R
~i{{\hat {\slash p'}+m_n} \over {\hat p'^2-m_n^2}} \ (-ie\gamma_\nu)P_R \, 
\nonumber \\  
 &\times &  i{{\hat {\slash p}+m_n} \over {\hat p^2-m_n^2}} \ (im_\mu P_L) 
\ ~i{{\hat {\slash p}+m_n} \over {\hat p^2-m_n^2}} \ ie P_R \ u(p) 
\ {1\over {k^2-m_A^{(n)~2}}} +h. c. \,,
\end{eqnarray}
As before, 
the amplitude where the mass insertion comes after the photon emission can 
be easily obtained by interchanging the ordering in the resulting final 
amplitude expression. Also, as before, a short analysis shows that graphs with 
external insertions or those involving a $W_5$ or a Higgs field lead to terms 
which are subleading in $m_\mu/M_{KK}$ or $m_\mu/M_W$. 
We thus obtain the total contribution 
to $g-2$ from a given KK level of a neutral gauge boson by adding 
the two expressions above and 
performing the momentum integrations; we find 
\begin{equation}
(g-2)_n={-3e^2g_Lg_Rm_\mu^2\over {16\pi^2 m_n^2}}\int^{1}_0 dx \int^{1-x}_0dy 
{6x^2+12xy-11x\over{1+\epsilon_n (x+y)}}\,,
\end{equation}
where we have defined $m_A^{(n)2}=m_n^2(1+\epsilon_n)$; note that 
$\epsilon_n=0$ in the case of photons. To go further we must sum over both 
the the photon and Z towers; using 
$\sum_1^\infty {1\over {n^2}}={\pi^2\over {6}}$ and 
$\sum_1^\infty {1\over {n^4}}={\pi^4\over {90}}$ we obtain the final numerical 
result 
\begin{equation}
(g-2)_{UED}=-439\cdot 10^{-11}(M_Z^2/M_c^2)[1-0.23(M_Z^2/M_c^2)+...]\,
\end{equation}
where we have neglected higher order terms in the ratio $M_Z^2/M_c^2$. For 
$M_c=300$ GeV, the smallest possible value, this gives $(g-2)_{UED}\simeq 
-40\cdot 10^{-11}$ which is only one quarter as large as the SM electroweak 
contribution.  
This is too small to make much of an impact on the potential difference between 
the experimental data and the SM prediction{\cite {bnl,marciano}}. Thus we 
conclude that $g-2$ does not yet provide any useful constraint on the UED 
scenario{\cite {uednew}}.

\section{Collider Production}

Due to the conservation of KK number at tree-level, KK excitations of the SM 
fields must be pair-produced at colliders. At $\gamma \gamma$ and 
lepton colliders the production 
cross sections for all the kinematically accessible KK states will very  
roughly be of order 100 fb $(1~TeV/\sqrt s)^2$ which yields respectable event 
rates 
for luminosities in the $100-500 ~fb^{-1}$ range. A sample of relevant cross 
sections at both $\gamma\gamma$ and $e^+e^-$ colliders are shown in Fig. 2.   
In the case of $\gamma \gamma$ collisions we have chosen the process 
$\gamma \gamma \to W^{+(1)}W^{-(1)}$ as it the process which 
has the largest cross section for 
the production of the first KK state. Similarly, gauge boson pair production 
in $e^+e^-$ collisions naturally leads to a large cross section. Clearly, 
such states once produced would not be easily 
missed for masses up to close the kinematic limit of the machine independently 
of how they decayed or if they were stable. To directly probe heavier masses 
we must turn to hadron colliders.

\vspace*{-0.5cm}
\nn
\begin{figure}[htbp]
\centerline{
\psfig{figure=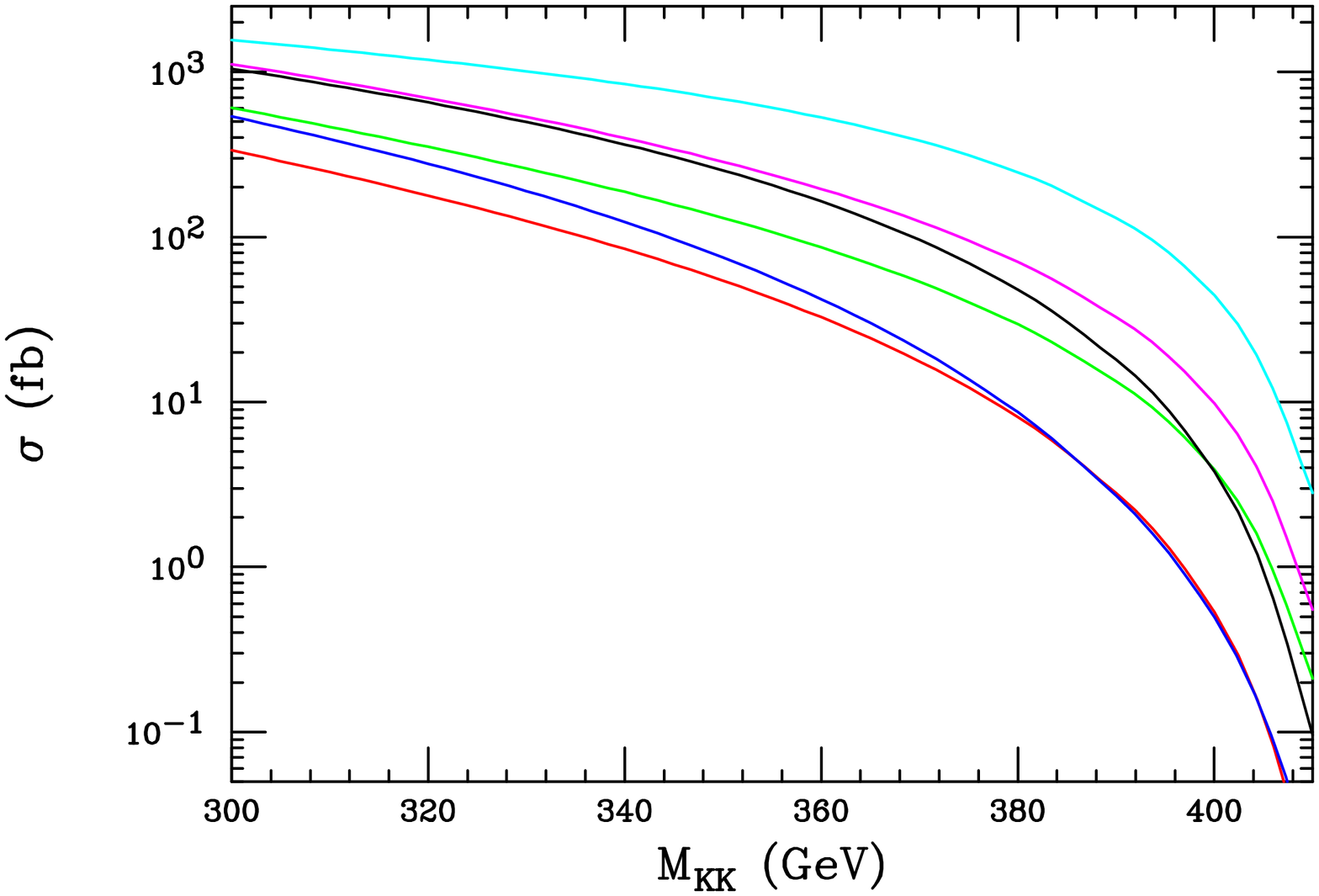,height=8.cm,width=8cm,angle=90}}
\vspace*{0.25cm}
\centerline{
\psfig{figure=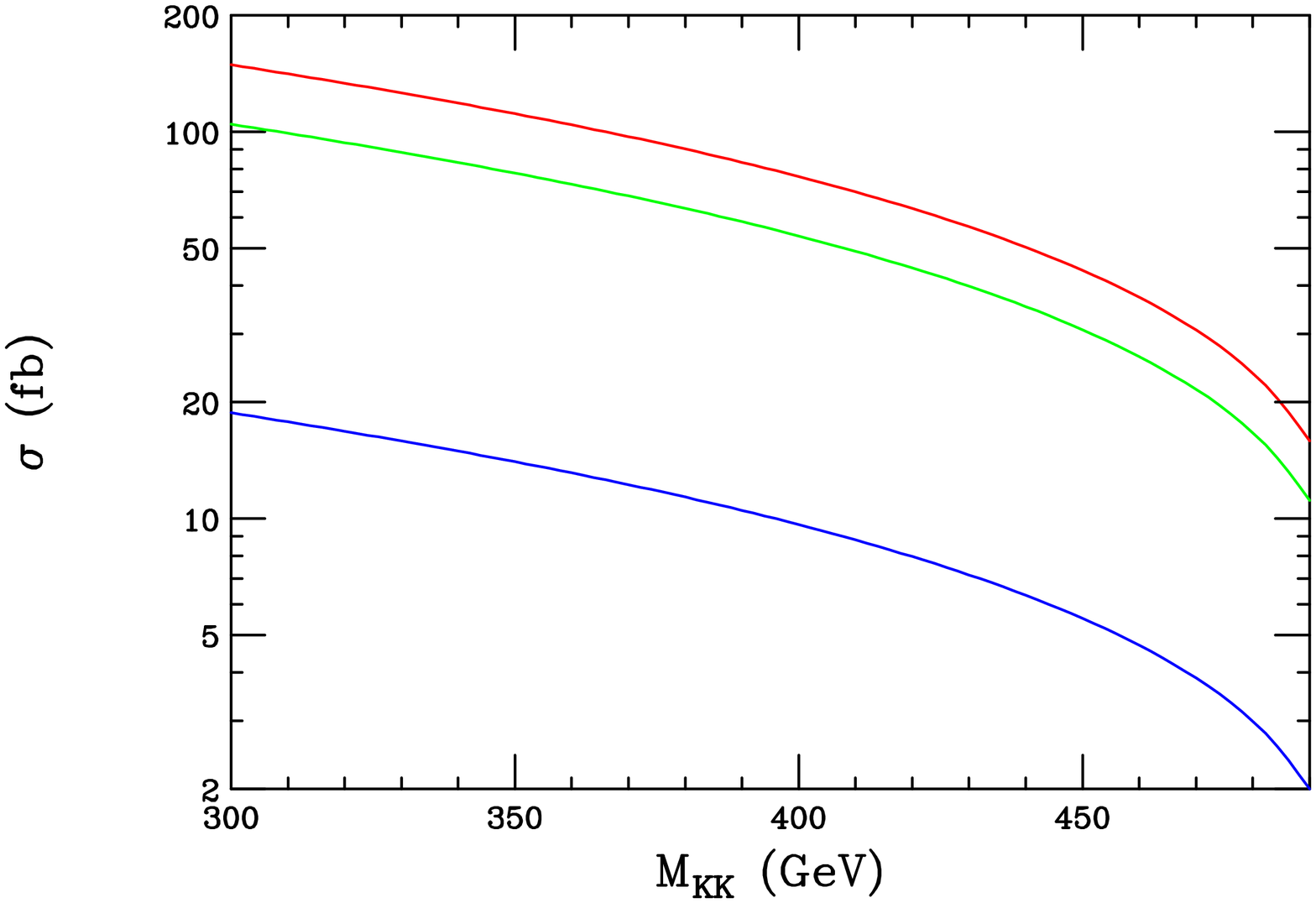,height=8.cm,width=8cm,angle=90}
\psfig{figure=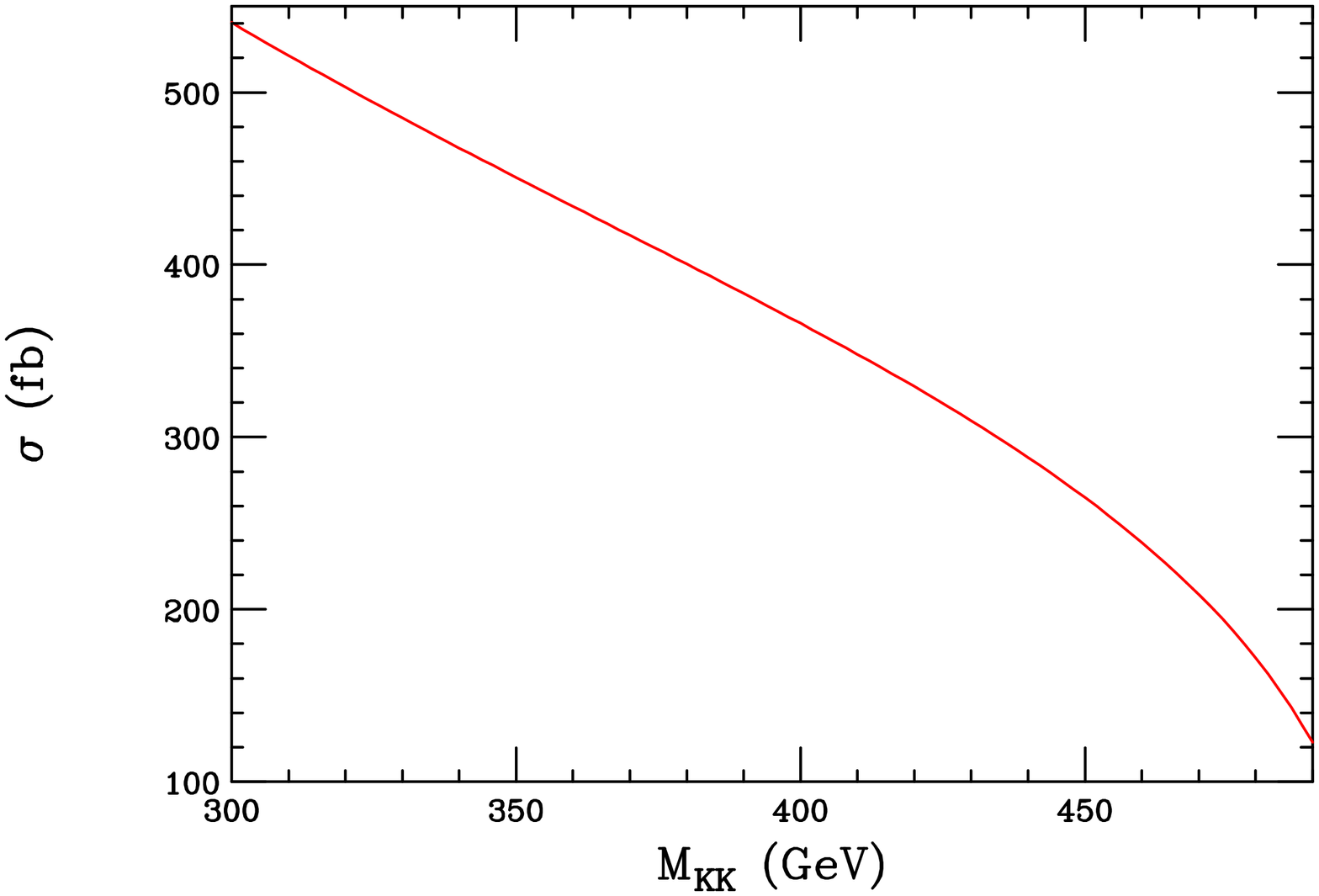,height=8.cm,width=8cm,angle=90}}
\vspace*{0.25cm}
\caption{Cross section for $\gamma\gamma \to W^{(1)+}W^{(1)-}$ (top panel) for 
different electron and laser polarizations for $\sqrt s_{ee}=1$ TeV. Cross 
section for $e^+e^- \to W^{(1)}W^{(1)}$ (lower right panel) for $\sqrt s=1$ 
TeV. Cross sections for (top to bottom, lower left panel) $e^+e^- \to 
2\gamma^{(1)}$, $Z^{(1)}\gamma^{(1)}$ and $2Z^{(1)}$ for $\sqrt s=1$ TeV.}
\end{figure}

Since both QCD and electroweak exchanges can lead to KK pair production at 
hadron colliders there are three classes of basic processes to consider. 
Clearly the states with color 
quantum numbers will have the largest cross sections at hadron machines and 
there are a number of processes which can contribute to their production at 
order $\alpha_s^2${\cite {duane}} several of which we list below:
\begin{eqnarray}
(i)~~gg &\to & g^{(1)}g^{(1)} \, \nonumber \\
(ii)~~qq'&\to & q^{(1)}q'^{(1)} \, \nonumber \\
(iii)~~gg+q\bar q &\to & q'^{(1)}\bar q'^{(1)} \, \nonumber \\
(iv)~~qq &\to & q^{(1)}q^{(1)} \, \nonumber \\
(v)~~q\bar q &\to & q^{(1)}q^{(1)} \,,
\end{eqnarray}
where the primes are present to denote flavor differences. 
Fig. 3 shows the cross sections for these five processes at both the 
$\sqrt s=2$ TeV Tevatron and the LHC summed over flavors. It is clear that 
the during the Tevatron Run II we should expect a reasonable yield of these 
KK particles for masses below $\simeq 600$ GeV if integrated luminosities in 
the range of 10-20 $fb^{-1}$ are obtained. Other processes that we have not 
considered may be able to slightly increase this reach. 
For larger masses we must turn to 
the LHC where we see that significant event rates should be obtainable for KK 
masses up to $\simeq 3$ TeV or so for an integrated luminosity of 100 
$fb^{-1}$. As one might expect we see that the most important QCD processes 
for the production of KK states are different at the two colliders.

\vspace*{-0.5cm}
\nn
\begin{figure}[htbp]
\centerline{
\psfig{figure=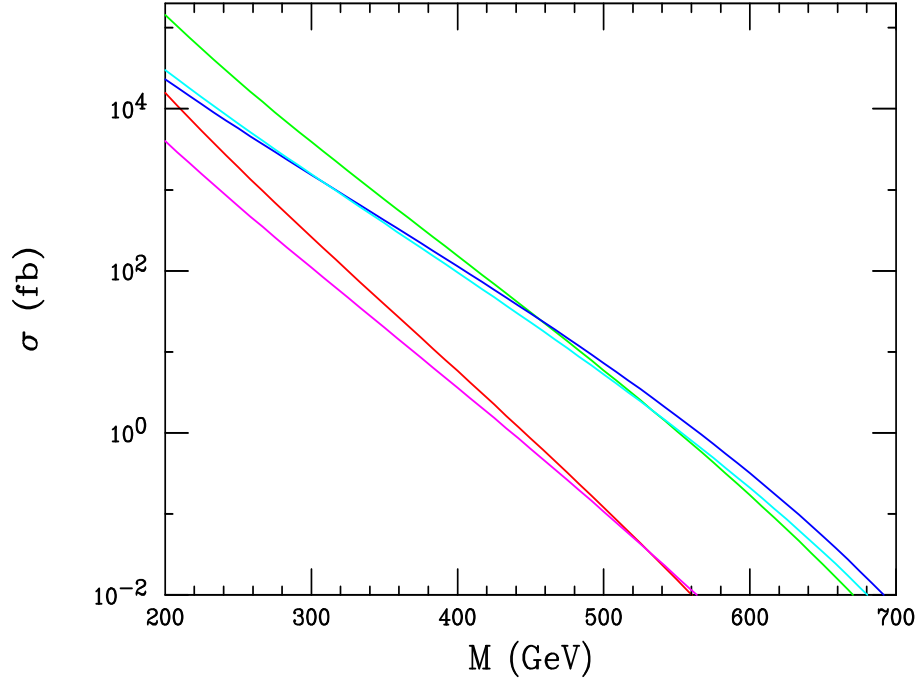,height=9cm,width=12cm,angle=90}}
\vspace*{15mm}
\centerline{
\psfig{figure=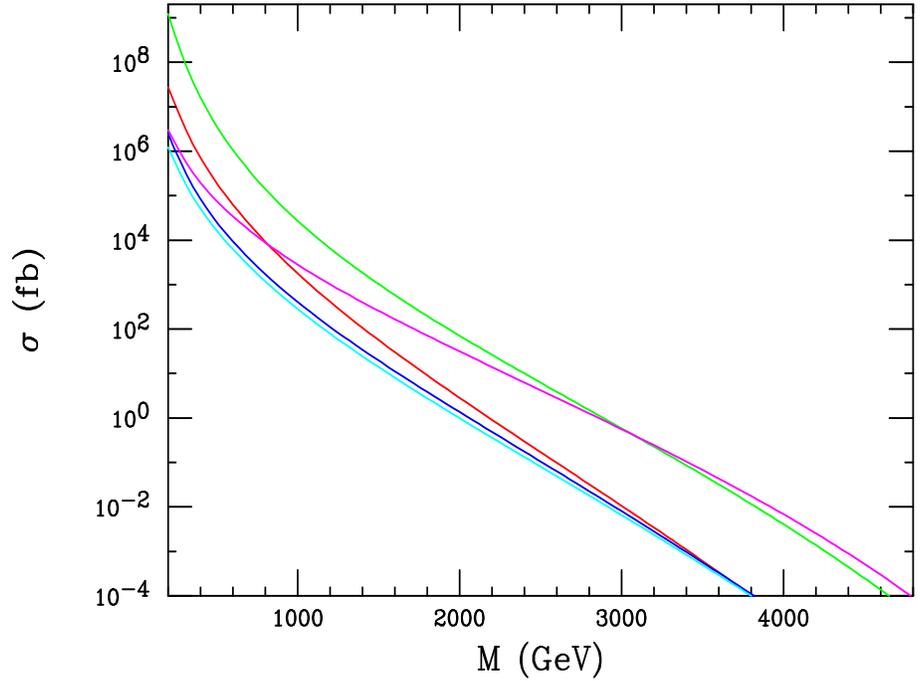,height=9cm,width=12cm,angle=90}}
\caption{Cross section for the pair production of the lightest colored KK 
states at the $\sqrt s=2$ TeV Tevatron(top) and the LHC(bottom). In the top 
panel, from top to bottom on the left-hand side, the curves correspond to the 
processes $ii,v,iii,i$ and $iv$, respectively. In the bottom 
panel, from top to bottom on the left-hand side, the curves correspond to the 
processes $ii,i,iv,iii$ and $v$, respectively. Antiquark contributions are 
included in reactions $ii$ and $iv$.}
\label{fun1}
\end{figure}
\vspace*{0.4mm}

The real signature of the UED scenario is that {\it all} of the SM fields 
have KK excitations. Thus we will also want to observe the production of the 
SM color singlet states. 
Of course color singlet states can also be produced with the largest cross 
sections being for associated production with a colored state at order 
$\alpha \alpha_s$; these rates are of course smaller than for pairs of colored 
particles as can be seen in Fig.~4. Here we see reasonable rates are obtained 
for KK masses in excess of $\simeq 1.8$ TeV or so. 

\vspace*{-0.5cm}
\nn
\begin{figure}[htbp]
\centerline{
\psfig{figure=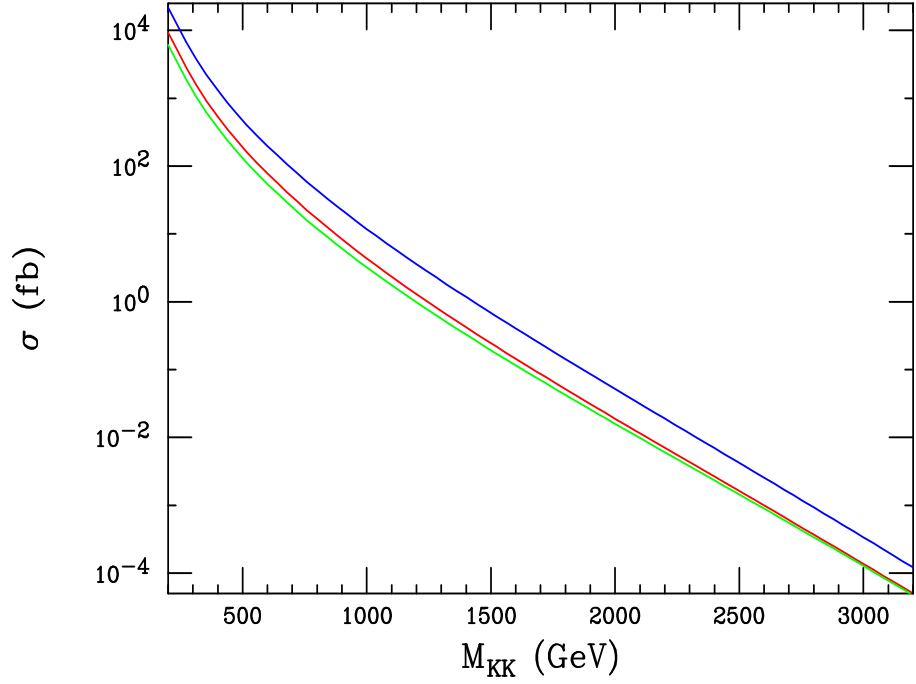,height=9cm,width=12cm,angle=90}}
\vspace*{15mm}
\centerline{
\psfig{figure=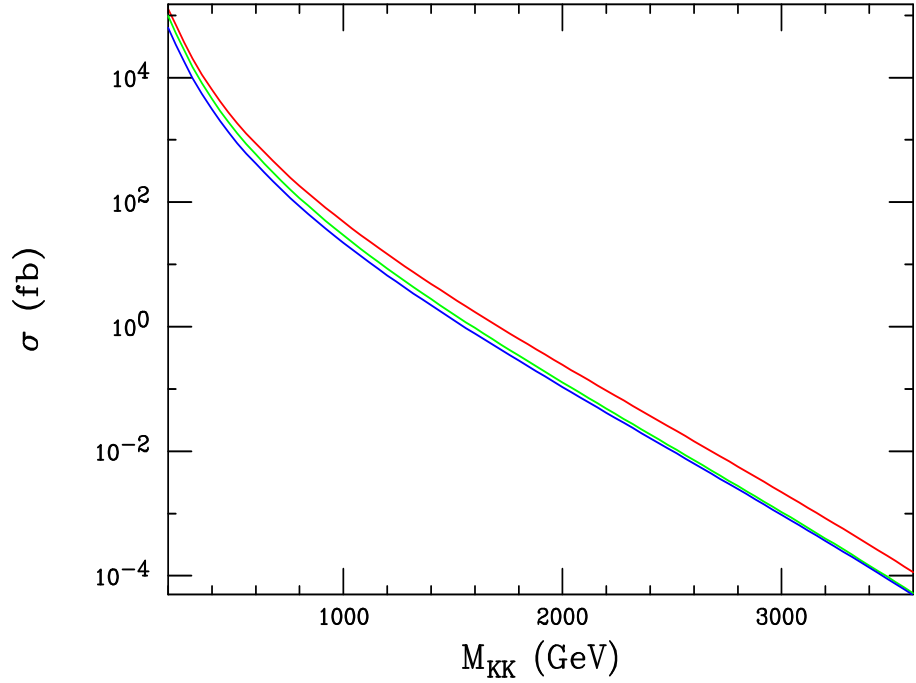,height=9cm,width=12cm,angle=90}}
\caption{Cross sections for the associated production of the lightest color 
singlet KK states at the LHC: in the top panel, from top to bottom, for 
$g^{(1)}W^{(1)\pm},g^{(1)}Z^{(1)}$ and $g^{(1)}\gamma^{(1)}$ final states; in 
the lower panel, from top to bottom, for $q^{(1)}W^{(1)\pm},q^{(1)}Z^{(1)}$ 
and $q^{(1)}\gamma^{(1)}$ final states. Anti-quark and contributions are 
included.}
\label{fun2}
\end{figure}
\vspace*{0.4mm}

Lastly, it is possible to pair produce color singlets via electroweak 
interactions which thus lead to cross sections of order $\alpha^2$. Due to the 
large center of mass energy of the LHC these cross sections can also lead to 
respectable production rates for KK masses as great as $\simeq 1.5$ TeV 
as can be seen from Fig.~5. 

\vspace*{-0.5cm}
\nn
\begin{figure}[htbp]
\centerline{
\psfig{figure=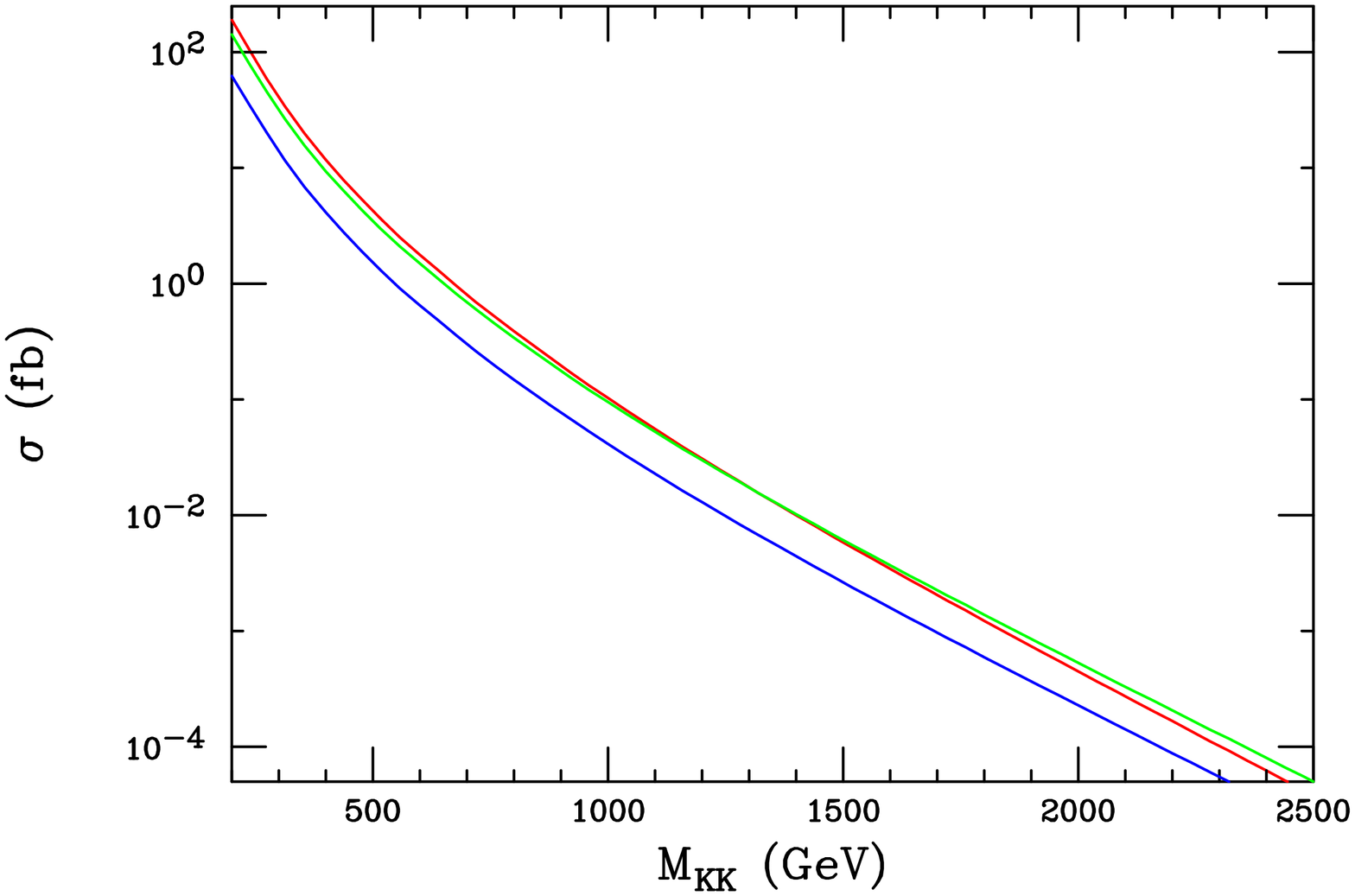,height=9cm,width=12cm,angle=90}}
\vspace*{15mm}
\centerline{
\psfig{figure=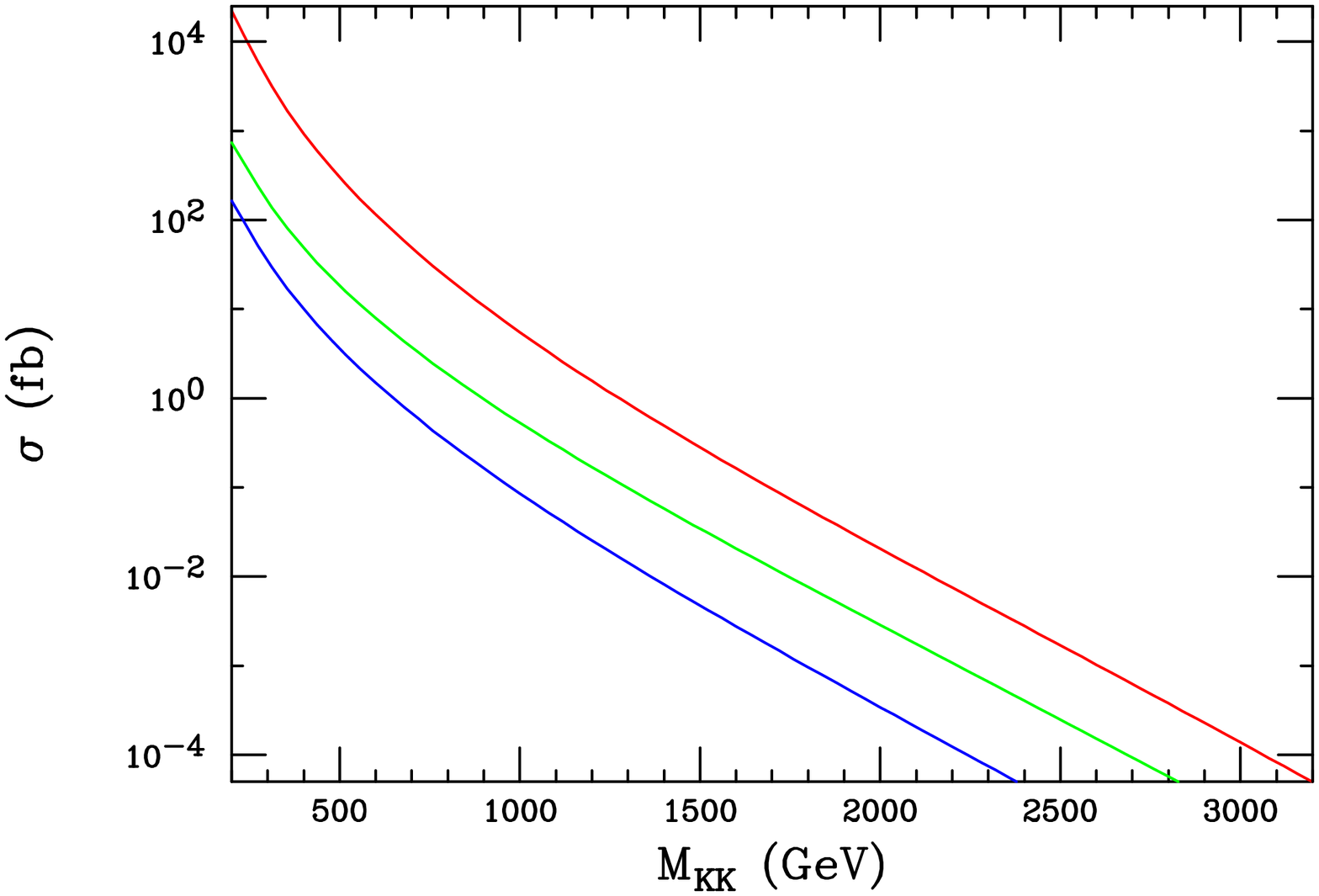,height=9cm,width=12cm,angle=90}}
\caption{Cross sections for the production of the lightest color singlet KK 
states at the LHC: in the top panel, from top to bottom, for $2Z^{1)},
\gamma^{(1)}Z^{(1)}$ and $2\gamma^{(1)}$ final states; in the bottom panel, 
from top to bottom, for $W^{(1)+}W^{(1)-},W^{(1)\pm}Z^{(1)}$ and 
$W^{(1)\pm}\gamma^{(1)}$ final states.}
\label{fun3}
\end{figure}
\vspace*{0.4mm}

It is clear from this analysis that the LHC will have 
a significant search reach 
for both colored and non-colored KK states provided that the production 
signatures are reasonably distinct. This is the subject for the next section.

\section{Collider Signatures}

When examining collider signatures for KK pair production in the UED there are 
two important questions to ask: ($i$) Are the lightest KK modes stable and 
($ii$) if they are unstable what are their decay modes? From the discussion 
above it is clear that without introducing any new physics the $n=1$ KK states 
{\it are} stable so we must consider this possibility when looking at 
production signatures.

In their paper ACD{\cite {ACD}} argue that cosmological constraints possibly 
suggest that KK states in the TeV mass range must be unstable on cosmological 
time scales. (Of course this does not mean that they would appear unstable 
on the time scale of a collider experiment in which case our discussion is 
the same as that above.) This would require the introduction of new physics 
beyond that contained in the original UED model. There are several possible 
scenarios for such new physics. Here we will discuss three possibilities in 
what follows, the first two of which 
were briefly mentioned by ACD{\cite {ACD}}.

Scenario I: The TeV$^{-1}$-scale UED model is embedded inside a thick brane 
in a higher $(\delta+4)$-dimensional space, with a compactification scale 
$R_G>>R_c$, 
in which gravity is allowed to propagate in a manner similar to the model of 
Arkani-Hamed, Dimopoulos and Dvali{\cite {nima}}. Since the graviton wave 
functions are normalized on a torus of volume $(2\pi R_G)^\delta$ while the KK 
states are normalized over $2\pi R_c$ the overlap of a KK zero mode with any 
even or odd KK tower state $n$ and a graviton will be non-zero. In a sense, 
the brane develops 
a transition form-factor analogous to that described in {\cite {derujula}}. 
This induces transitions of the form $KK(n=1)\to KK(n=0)+G$ where $G$ 
represents the graviton field and appears as missing energy in the collider 
detector. This means that production of a pair of KK excitations of, \eg,  
quarks or gluons would appear as two jets plus missing energy in the detector; 
the corresponding production of a KK excited pair of gauge bosons would appear 
as the pair production of the corresponding zero modes together with missing 
energy. We can express this form-factor simply as 
\begin{equation}
{\cal F}={\sqrt 2\over {\pi R_c}}\int_0^{\pi R_c}dy
e^{im_gy}(\cos ny/R_c,\sin ny/R_c)\,,
\end{equation}
for even and odd KK states, respectively, 
where $m_g$ is the graviton mass. Here we have assumed that the thick brane 
resides at $y_i=0$ for all $i\neq 1$. These integrals can be performed 
directly and we obtain the following expressions for the transition 
form-factors in the case where $n=1$:
\begin{equation}
{\cal F}^2_{even}(n=1)={4\over {\pi^2}}~{x^2\over {(1-x^2)^2}}~(1-\cos (\pi x))
\,,
\end{equation}
with ${\cal F}^2_{odd}(n=1)={\cal F}^2_{even}(n=1)/x^2$ where $x=m_gR_c$. 
Given these form-factors we can calculate the actual decay rate for
$KK(n=1)\to KK(n=0)+G$, where we now must 
sum up the graviton towers by following the 
analyses in Ref.{\cite {feyn}}; this result should be relatively independent 
of the spin of the original KK state. We find the total with to be given by 
\begin{equation}
\Gamma={(2\pi)^{\delta/2}~\mpl^2\over {\Gamma(\delta/2)~M_D^{2+\delta}}}
\int^{M_{KK}}_{R_G^{-1}} dm_g 
~m_g^{\delta-1}~\Gamma(m_g)~[{\cal F}(m_gR_c)]^2(n=1)
\,,
\end{equation}
where $\Gamma(m_g)$ is the width for the decay into a graviton of mass $m_g$, 
$M_D$ is the $\delta+4$-dimensional Planck scale, $\mpl$ is the conventional 
4-d reduced Planck scale, and $M_{KK}$ is the mass of the relevant decaying KK 
state. Performing the integration numerically we 
obtain the results shown in Fig.6. This figure shows that this mechanism 
provides for a very rapid decay over almost all of the parameter space. For 
light KK states with both $\delta$ and $M_D$ large the decay rate is suppressed 
and may lead to finite length charged tracks in the detector. (In particular 
the production of a charged KK state with a long lifetime would yield a 
kink-like track structure 
due to the decay to the graviton tower.) Although not a 
true two-body decay, Fig.6 also shows that the typical missing energy in the 
gravitational decay of a KK state will be close to half its mass, which is 
quite significant for such heavy objects. It is clear that events with such a 
large fraction of missing energy should be observable above background given 
sufficient event rates. These events will not be confused with SUSY signals 
since they occur in every possible channel. 

\nn
\begin{figure}[htbp]
\centerline{
\psfig{figure=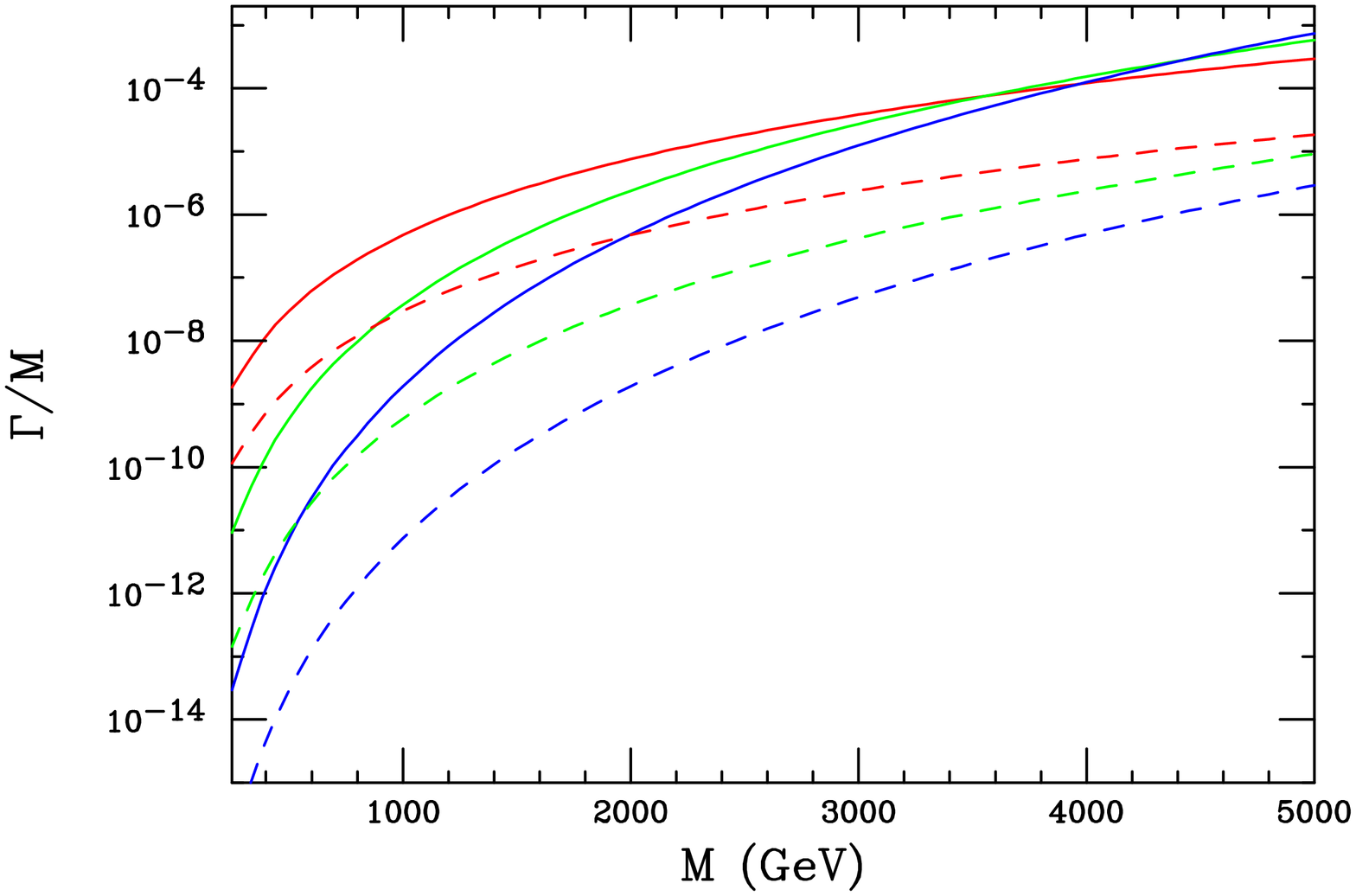,height=8.cm,width=8cm,angle=90}
\psfig{figure=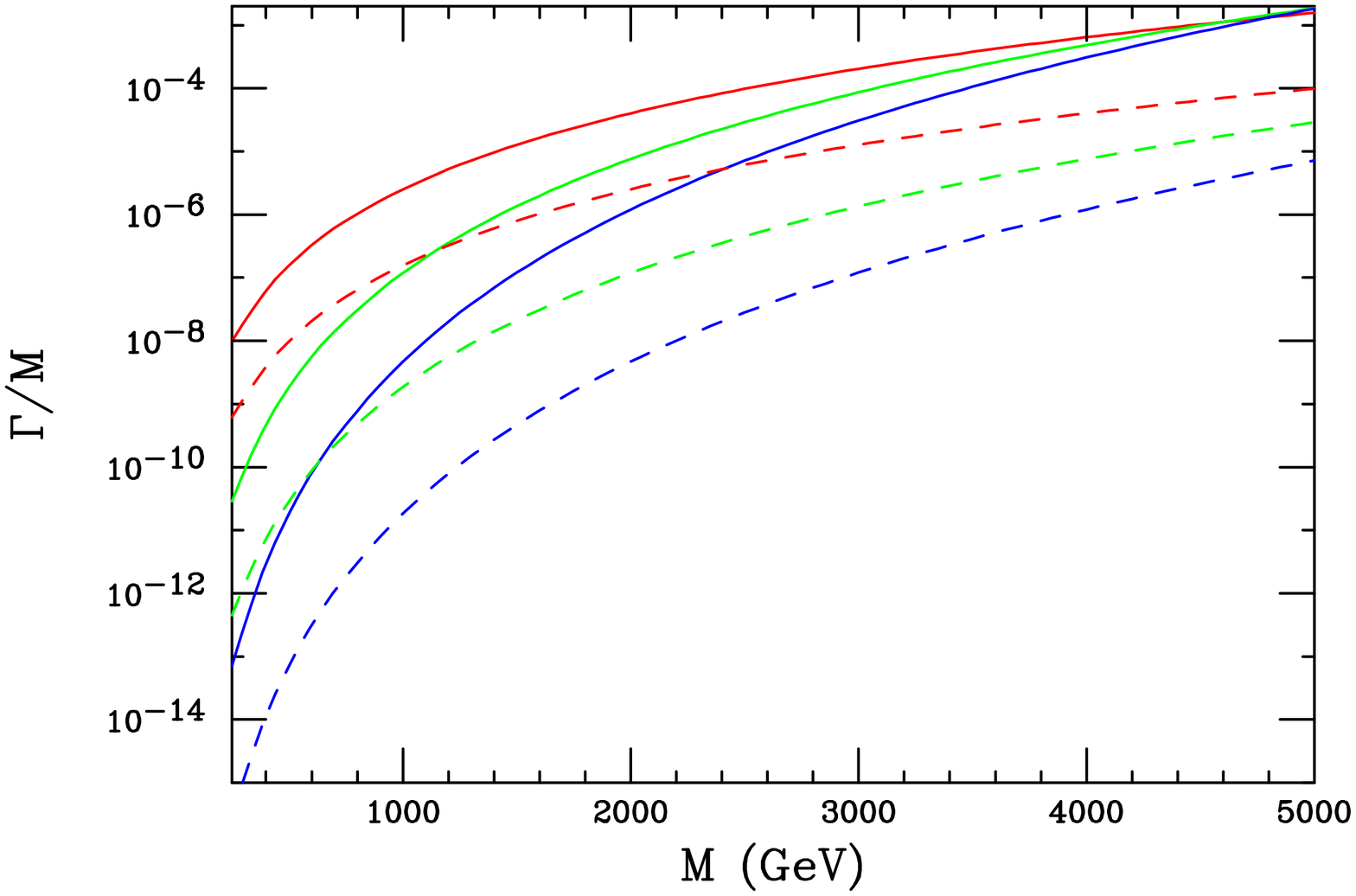,height=8.cm,width=8cm,angle=90}}
\vspace*{0.25cm}
\centerline{
\psfig{figure=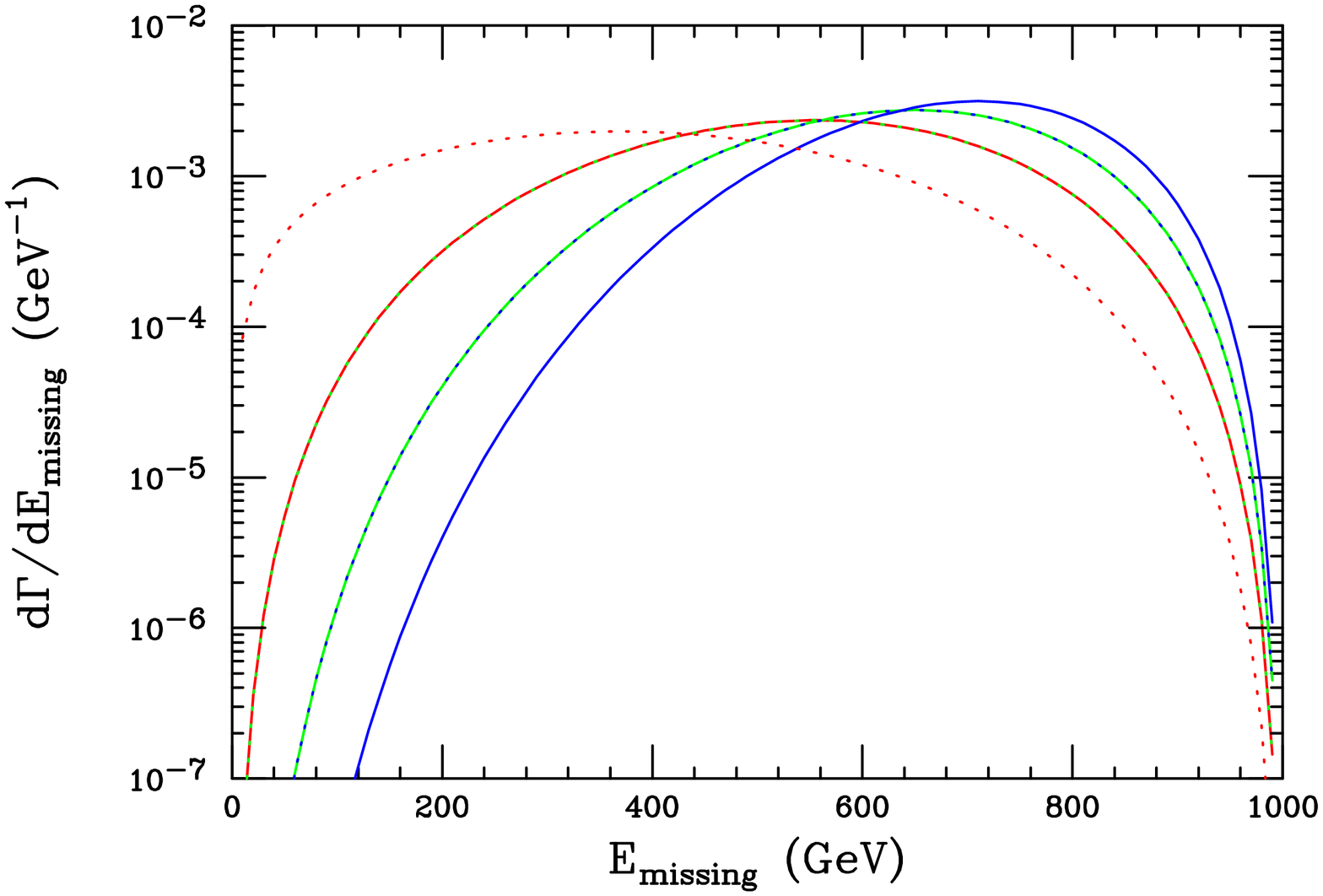,height=8.cm,width=8cm,angle=90}}
\vspace*{0.25cm}
\caption{Width for the decay of the first excited KK state (even-top left 
panel, 
odd-top right panel) into the corresponding zero mode and a graviton tower as a 
function of the mass of the KK state. The solid(dashed) lines are for 
$M_D=5(10)$ TeV and from top to bottom in each case the curves correspond to 
$\delta=2,4,6$, respectively. The lower panel shows the missing energy 
distribution for these decays for these same cases assuming a KK mass of 1 
TeV.}
\end{figure}

Scenario II: KK decays can be induced in the UED model by adding a `benign' 
brane at some $y=y_0$ which induces new interactions. By `benign' we mean 
that these new interactions only do what we need them to do and do not alter 
the basic properties of the UED model. The simplest form of such interactions 
are just the four dimensional variants of the terms in the the 5-d UED 
action. For example, one might add a term such as 
\begin{equation}
\int d^4x \int dy ~\delta(y-y_0) ~{\lambda\over {M_s}} \bar \psi \gamma^A 
{\cal D}_A \psi \,,
\end{equation}
where $\lambda$ is some Yukawa-like coupling and $M_s$ is some large scale.  
Note that the brane is placed at some arbitrary position $y=y_0$ and {\it not} 
at the fixed points where only even KK modes would be effected. These new 
interactions result in a mixing of all KK states both even and odd and, in 
particular, with the zero mode. Thus we end up inducing decays of the form 
KK$^{(1)}\to$ KK$^{(0)}$ KK$^{(0)}$. For KK fermions the decay into a fermion 
plus gauge boson zero mode is found to be given by
\begin{equation}
\Gamma(f^{(1)}\to f^{(0)}V^{(0)})={g_V^2\over {8\pi}} s_\phi^2 M_c \cdot N_c 
\cdot PS
\,,
\end{equation}
where $s_\phi$ is the induced mixing angle, $N_c$ is a color factor, $g_V$, 
the relevant gauge coupling and PS is the phase space for the decay. It is 
assumed that the mixing angle is sufficiently small that single production 
of KK states at colliders remains highly suppressed but is large enough for 
the KK state to decay in the detector. For $\lambda \simeq 0.1$ and $M_s 
\simeq$ a few $M_c$ this level of suppression is quite natural. (Numerically, 
it is clear that the KK state will decay inside the detector unless the mixing 
angle is very highly suppressed.) The resulting branching fractions can 
be found in Table 1 where we see numbers that are not too different than those 
for excited fermions in composite models with similar decay signatures. 
However, unlike excited SM fields, single production modes are highly 
suppressed. For KK excitations of the gauge 
bosons, their branching fractions into zero mode fermions will be identical to 
those of the corresponding SM fields apart from corrections due to phase 
space, \ie, the first excited $Z$ KK state can decay to $t\bar t$ while the 
SM $Z$ cannot.

\begin{table}
\centering
\begin{tabular}{|l|c|c|c|c|} \hline\hline
       &  $g$  &  $\gamma$  &  $Z$  &  $W$  \\ \hline \hline
 $e^{(1)}$&   0   &   41.0&     14.4&   44.6\\
 $\nu^{(1)}$&   0   &   0&     39.1&   60.9\\
 $u^{(1)}$&   89.8   &    2.3&      2.1&    5.7\\
 $d^{(1)}$&   90.9   &    0.6&      2.7&    5.8\\ \hline\hline
\end{tabular}
\caption{Individual branching fractions in per cent 
for the first excited fermion KK modes when KK level mixing occurs as in 
Scenario II.}
\end{table}

Scenario III: We can add a common bulk mass term to the fermion action, \ie, 
a term of the form 
$m(\bar DD+\bar SS)$; we chose a common mass for both simplicity and to avoid 
potentially dangerous flavor changing neutral currents. The largest influence 
of this new term is to modify the zero mode fermion wavefunction which is now 
no longer flat and takes the form $\sim e^{-m|y|}$
and thus remains $Z_2$-even. Clearly there is now a significant overlap in the 
5-d wavefunctions 
between pairs of fermion zero modes and any $Z_2$-even gauge KK mode which can  
be represented as another form-factor:

\begin{equation}
{\cal G}(x)={4x^2\over {4x^2+n^2}}~{1-(-1)^ne^{-2\pi x}\over {(1-e^{-2\pi x})}}
\,,
\end{equation}
where $x=mR_c$ and $n$ is the KK mode number. This form factor then describes 
the decay $G^{(n)}\to \bar f^{(0)}f^{(0)}$ where $G$ represents a generic 
KK gauge 
field. Similarly we can obtain a form-factor that describes the corresponding 
decay $f^{(n)}\to G^{(0)}f^{(0)}$ given by 

\begin{equation}
{\cal G}'(x)={2\over {\sqrt{\pi x}}}~{x^2\over {x^2+n^2}}~{1-(-1)^ne^{-\pi x}
\over {(1-e^{-2\pi x})^{1/2}}}\,,
\end{equation}
where $x$ is as above. It is clear that the decays of KK states in this 
scenario will be essentially identical to Scenario II above although they are 
generated by a completely different kind of physics. 
Fig. 7 shows the shape of these two form-factors as 
a function of the parameter $x$. The natural question to ask at this point 
is `what is the value of $m$ relative to $M_c$?'. It seems natural to imagine 
that the bulk mass would be of order the compactification scale, the only 
natural scale in the action, which would 
imply that $x\sim 1$ so that large form-factors would be obtained. While this 
scenario works extremely well for the decay of $Z_2$-even states it does not 
work at all for the case of $Z_2$-odd states.

\vspace*{-0.5cm}
\nn
\begin{figure}[htbp]
\centerline{
\psfig{figure=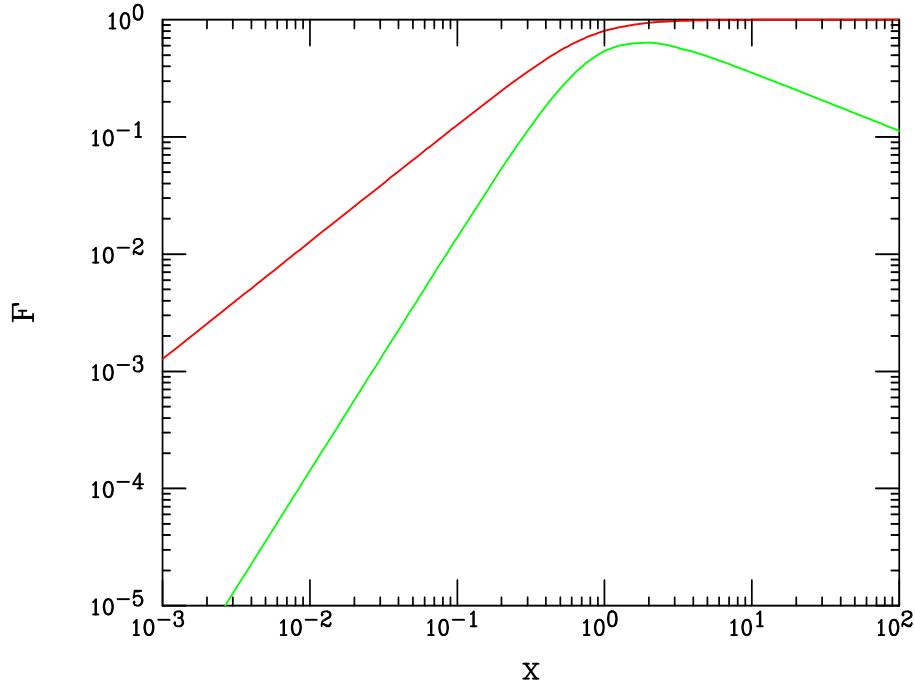,height=9cm,width=12cm,angle=90}}
\vspace*{0.5cm}
\caption[*]{The form factors ${\cal G}$(upper curve) and ${\cal G}$'(lower 
curve) as discussed in the text for the case $n=1$.}
\label{lims}
\end{figure}

\section{Summary and Conclusions}

In this paper we have begun a detailed examination of the predictions of the 
Universal Extra Dimensions model for future colliders. Since it is known from 
the detailed analysis of Appelquist, Cheng and Dobrescu and the subsequent 
work by Agashe, Deshpande and Wu that the compactification scale, $M_c$ in 
this model 
can be as low as 350 GeV, we first examined the contribution to $g-2$ in this 
model arising from loops of KK states. Although this contribution, of order 
$-40\times 10^{-11}$, may eventually be probed by the Brookhaven $g-2$ 
experiment, no new bounds on $M_c$ are at present obtainable. Next we turned 
to the production of the lightest KK states at lepton, $\gamma \gamma$ and 
hadron colliders. Here it was necessary to be reminded that due to tree-level 
conservation of KK number in the UED model it is necessary to pair-produce 
KK states. Since indirect searches for such states give rather poor reaches 
direct 
searches are of greater importance in this model than in the other cases 
discussed in the introduction. Thus to 
obtain interesting search reaches requires a hadron collider such as the 
Tevatron or LHC. Based on counting events we expect the reach st the Tevatron 
Run II (LHC) for KK states to be $\simeq 600(3000)$ GeV. Within the UED 
model itself these lightest KK states are stable even when loop corrections 
are included unless new interactions are introduced from elsewhere. If these 
states are indeed stable, the production of a large number of heavy stable 
charged
particles would not be missed at either collider. It is more likely, however, 
that new physics does indeed enter rendering the KK modes unstable. In this 
paper we have examined three new physics scenarios that induce finite KK 
lifetimes and compared their decay signatures. 

In the first case the UED model 
was embedded in a thick brane in a larger n-dimensional space 
in which gravity was 
free to propagate. Due to the difference in sizes over which the various 
fields are normalized, excited KK states can now decay to zero mode SM fields 
through the emission of gravitons whose rate is 
controlled by a geometric form-factor 
like function. At colliders this would appear as the 
production of pairs of SM states in association with a large amount of 
missing energy from the two 
towers of gravitons. In a second scenario, a `benign' 
brane is introduced somewhere between the fixed points on which a set of 
non-renormalizable interactions occur. These interactions then induce mixing 
amongst the various KK levels violating KK number conservation and allowing 
excited KK modes to decay to SM fields. The branching fractions of all of the 
fermionic KK states were calculated while those of the gauge KK states are 
found to be essentially the same as the corresponding SM fields 
apart from phase space effects. 
In the last scenario we consider the possibility that the 5-dimensional 
fermion fields obtain a common bulk mass; a common bulk mass was assumed both 
on the basis of simplicity and to avoid any potentially dangerous FCNC. 
This modifies the wave-function of 
the zero modes so that a finite overlap exists with higher modes. This then 
allows the decay of KK states through another set of form-factors that arise 
from these wave function overlaps. Unfortunately these form-factors vanish for 
$Z_2$-odd KK excited states due to $Z_2$ parity conservation and thus these 
states will remain stable. In this scenario the decay signatures are 
found to be similar to those of the previous case. 

Clearly, independent of whether the first excited KK modes are stable or decay 
through one of the above mechanisms, if the UED framework is at all correct 
future colliders will yield exciting signals of new physics associated with 
extra dimensions.

\noindent{\Large\bf Acknowledgements}

The author would like to thank H.-C Cheng and B. Dobrescu for discussions 
during the early stages of this work and the CERN Theory Division for its 
hospitality. The author would also like to thank H. Davoudiasl and J.L. Hewett 
for general discussions on models with extra dimensions.

\newpage

%
\def\MPL #1 #2 #3 {Mod. Phys. Lett. {\bf#1},\ #2 (#3)}
\def\NPB #1 #2 #3 {Nucl. Phys. {\bf#1},\ #2 (#3)}
\def\PLB #1 #2 #3 {Phys. Lett. {\bf#1},\ #2 (#3)}
\def\PR #1 #2 #3 {Phys. Rep. {\bf#1},\ #2 (#3)}
\def\PRD #1 #2 #3 {Phys. Rev. {\bf#1},\ #2 (#3)}
\def\PRL #1 #2 #3 {Phys. Rev. Lett. {\bf#1},\ #2 (#3)}
\def\RMP #1 #2 #3 {Rev. Mod. Phys. {\bf#1},\ #2 (#3)}
\def\NIM #1 #2 #3 {Nuc. Inst. Meth. {\bf#1},\ #2 (#3)}
\def\ZPC #1 #2 #3 {Z. Phys. {\bf#1},\ #2 (#3)}
\def\EJPC #1 #2 #3 {E. Phys. J. {\bf#1},\ #2 (#3)}
\def\IJMP #1 #2 #3 {Int. J. Mod. Phys. {\bf#1},\ #2 (#3)}
\def\JHEP #1 #2 #3 {J. High En. Phys. {\bf#1},\ #2 (#3)}


\begin{thebibliography}{99}
\bibitem{strumia} 
The author would like to thank A. Strumia for this important observation.
\end{thebibliography}
\end{document}